\documentclass[aps,prl,preprint,showpacs]{revtex4}
\usepackage{bm}
\usepackage{graphicx}
\begin{document}
\title{Mesoscopic Theory of Critical Fluctuations in Isolated Granular Gases}
\author{J. Javier Brey}
\email{brey@us.es}
\author{A. Dom\'{\i}nguez}
\author{M.I. Garc\'{\i}a de Soria}
\author{P. Maynar}
\affiliation{F\'{\i}sica Te\'{o}rica, Universidad de Sevilla,
Apdo.\ de Correos 1065, E-41080 Sevilla, Spain}

\date{today}

\begin{abstract}
Fluctuating hydrodynamics is used to describe the total energy
fluctuations of a freely evolving gas of inelastic hard spheres
near the threshold of the clustering instability. They are shown
to be governed by vorticity fluctuations only, that also lead to a
renormalization of the average total energy. The theory predicts a
power-law divergent behavior of the scaled second moment of the
fluctuations, and a scaling property of their probability
distribution, both in agreement with simulations results. A more
quantitative comparison between theory and simulation for the
critical amplitudes and the form of the scaling function is also
carried out.

\end{abstract}


\maketitle

Granular gases are assemblies of macroscopic particles evolving
independently between inelastic collisions \cite{JNyB96}. The
methods of non-equilibrium statistical mechanics, kinetic theory,
and hydrodynamics have been successfully extended to describe the
observed macroscopic behavior and also, although in a much more
limited form, the fluctuations around it \cite{Go03}. The lack of
energy conservation makes them to behave quite differently from
molecular fluids. A simple widely used model for granular gases
consists of smooth inelastic hard spheres (IHS), with momentum
conserving dynamics. Inelasticity is characterized by means of a
constant coefficient of normal restitution $\alpha$.

Recently, molecular dynamics (MD) simulation results have been
reported  for the total energy fluctuations of a two-dimensional
freely evolving IHS gas, near the threshold of the clustering
instability \cite{BGMyR05}.  The dimensionless second moment was
found to exhibit a power-law divergent behavior with the distance
to the instability. Also, the scaled cooling rate was found to
tend to zero according to a power law, although in a weak way.
Besides, the distribution function for the energy fluctuations,
when properly scaled, turned out to be independent of the
parameters defining the system. This was associated with a scaling
property of the distribution. Quite remarkably, the scaling
function was very well fitted by the same expression as several
equilibrium and non-equilibrium molecular systems
\cite{BHyP98,BChFyal00}.

The main goal of the present Letter is to provide an explanation
for the above results on the basis of fluctuating hydrodynamics
\cite{LyL59}, showing that it gives an accurate description of the
fluctuations occurring in a granular fluid in the threshold of the
clustering instability.

Consider an isolated system of $N$ IHS of mass $m$ and diameter
$\sigma$. The total (kinetic) energy $\tilde{E}$ of the system can
be expressed in the form \cite{ByE98}
\begin{equation}
\label{1} \tilde{E}(t)= \frac{1}{2} \int \!d{\bm r}\, \left[ d \tilde{n}({\bm
r},t)\tilde{T}({\bm r},t)+ m \tilde{n}({\bm r},t)
\tilde{u}^{2}({\bm r},t) \right],
\end{equation}
where $d$ is the dimension of the system, $\tilde{n}({\bm r},t)$
the number density field, $\tilde{T}({\bm r},t)$ the temperature
field, and $\tilde{\bm u}({\bm r},t)$ the flow field. The tildes
indicate that all the quantities are understood as fluctuating
variables. The above expression can be justified by identifying
the definitions of the microscopic densities with their
fluctuating values. In the following, systems in the {\em
homogeneous cooling state} (HCS) will be considered. At a
macroscopic level, this state is characterized by a constant
uniform density $n_{H}$, a vanishing flow field ${\bm u}_{H}=0$,
and a uniform time-dependent temperature obeying the law
\cite{Ha83} $\partial_{t}T_{H}(t)=-\zeta_{H}(T_{H})T_{H}(t)$,
where $\zeta_{H} \propto T_{H}(t)^{1/2}$ is the cooling rate.
Moreover, we will restrict ourselves to the region in which the
amplitudes of the fluctuations of the fields around their HCS
values remain small on the average. Then retaining up to quadratic
order in the deviations, Eq. (\ref{1}) yields
\begin{eqnarray}
\label{2} \delta \tilde{E}(t) & \equiv & \tilde{E}(t)-E_{H}(t)
\nonumber \\
&=& \frac{1}{2} \int d{\bm r}\, \left[ d n_{H} \delta
\tilde{T}({\bm r},t)+  d \delta \tilde{n}({\bm r},t) \delta
\tilde{T}
({\bm r},t) \right. \nonumber \\
& & + \left. m n_{H} |\delta \tilde {\bm u} ({\bm r},t)|^{2}
\right] .
\end{eqnarray}
Here, $E_{H}(t)=d N T_{H}(t)/2$, $\delta \tilde{n}({\bm
r},t)=\tilde{n}({\bm r},t)-n_{H}$, $\delta \tilde{u}({\bm
r},t)=\tilde{u}({\bm r},t)$, and $\delta \tilde{T}({\bm r},t)=
\tilde{T}({\bm r},t)-T_{H}(t)$. It is now convenient to introduce
dimensionless position, ${\bm l}$, and  time, $s$, scales as ${\bm
l}={\bm r}/{l_{0}}$ and $ds= v_{H}(t)dt /l_{0}$, respectively,
where $v_{H} \equiv \left[ 2T_{H}(t)/m \right]^{1/2}$ is the
termal velocity and $l_{0} \equiv (n_{H}\sigma^{d-1})^{-1}$ is
proportional to the mean free path. Moreover, dimensionless fields
are defined by $\rho({\bm l},s)= \delta \tilde{n}({\bm
r},t)/n_{H}$, ${\bm \omega}({\bm l},s)=\delta \tilde{\bm u}({\bm
r},t)/v_{H}(t)$, and $\theta ({\bm l},s)= \delta \tilde{T}({\bm
r},t)/T_{H}(t)$. Then, Eq.\ (\ref{2}) takes the form
\begin{equation}
\label{3} \epsilon(s)=\frac{\theta_{\bm 0}(s)}{V}+\frac{1}{V^{2}}
\sum_{\bm k} \left[ \rho_{\bm k}(s) \theta_{- {\bm k}} (s)
+\frac{2}{d} |{\bm \omega}_{\bm k}(s)|^{2} \right],
\end{equation}
where $\epsilon (s) \equiv \delta \tilde{E}(s)/E_{H}(s)$,
$V=L^{d}$ is the volume of the system in the new units, and the
Fourier transforms of the fields have been introduced. It is
assumed that after a time of the order of the mean free time
between collisions, the system reaches a hydrodynamic regime in
which all the energy of the system is stored in the hydrodynamic
modes. In this regime, the hydrodynamic fields are expected to be
described at a mesoscopic level by fluctuating hydrodynamic
equations. Here, they will be assumed to be linear Langevin
equations obtained by linearizing the Navier-Stokes equations for
a granular gas around the HCS. Moreover, the assumption is made
that the noise terms are defined by the same properties as for
molecular, elastic gases \cite{vNEByO97}. This is not expected to
be true, except in the nearly elastic limit, i.e. when the
coefficient of normal restitution $\alpha$ is very close to unity.
Consequently, the theory will be restricted in the following to
this limit.
  Thus, the transversal flow field or vorticity field, ${\bm \omega}_{{\bm k}
\perp}$, obeys the equation \cite{vNEByO97,LyL59}
\begin{equation}
\label{4} \left( \partial_{s}-\zeta^{*} /2+\eta^{*} k^{2} \right)
{\bm \omega}_{{\bm k} \perp}(s) = {\bm \xi}_{{\bm k} \perp}(s).
\end{equation}
In this expression, $\zeta^{*}=\zeta_{H}[T_{H}(t)] l_{0}/v_{H}(t)$
and $\eta^{*}=\eta_{H}[T_{H}(t)]/m n_{H}l_{0}v_{H}(t)$, $\eta_{H}$
being the shear viscosity. The random noise term ${\bm \xi}_{{\bf
k} \perp}(s)$ is Gaussian, with a correlation
\begin{equation}
\label{5} \langle {\bm \xi}_{{\bm k} \perp}(s)  {\bm \xi}_{{\bm
k}^{\prime} \perp}(s^{\prime}) \rangle=\frac{V^{2}}{N} \delta
(s-s^{\prime}) \delta_{{\bm k},-{\bm k}^{\prime}} \eta^{*} k^{2}
\sf{I},
\end{equation}
${\sf I}$ being the unit tensor of dimension $d-1$ in the subspace
perpendicular to ${\bm k}$, and the angular brackets denoting
average over the noise realizations. A main advantage of using the
scaled variables is that the coefficients in Eq.\, (\ref{4}) and
the strength of the noise become time-independent, contrary to
what happens in the original variables. The equation shows that
${\bm \omega}_{{\bm k} \perp}$ grows in time for those values of
${\bm k}$ such that $\lambda_{\perp}({\bm k}) \equiv
\zeta^{*}/2-\eta^{*}k^{2} >0$. Although this does not imply by
itself that the HCS is linearly unstable, due to the
time-dependent scaling of the velocity introduced above,
simulation results and nonlinear analytical analysis of the
Navier-Stokes equations have shown that this growth is the origin
of the clustering instability \cite{GyZ93,BRyC99}. The minimum
value of $k$ for a system of linear extent $L$, measured in the
$l$-scale, is $k_{min}=2 \pi/L$. Then, for given values of the
other parameters, the system becomes unstable if $L>L_{c}$, with
$L_{c}= 2 \pi ( 2 \eta^{*} / \zeta^{*} )^{1/2}$. For $L<L_{c}$,
the HCS is stable and the long time solution of Eq. (\ref{4}) is
\begin{equation}
\label{6} {\bm \omega}_{{\bm k} \perp}(s)= \int_{- \infty}^{s}
ds^{\prime}\, e^{(s-s^{\prime})\lambda_{\perp}(k)} {\bm \xi}_{{\bm
k} \perp} (s^{\prime}).
\end{equation}
From this expression, it is easily obtained
\begin{equation}
\label{7} \langle {\bm \omega}_{{\bm k} \perp}(s) {\bm
\omega}_{{\bm k}^{\prime} \perp}(s^{\prime}) \rangle =-
\frac{V^{2} \eta^{*} k^{2}}{2N
\lambda_{\perp}(k)}e^{(s-s^{\prime}) \lambda_{\perp}(k)}
\delta_{{\bm k},-{\bm k}^{\prime}} {\sf I},
\end{equation}
for $s \geq s^{\prime} \gg1 $. This shows that as $L$ tends to
$L_{c}$ from below, the amplitudes of the fluctuations of the
transversal components of the velocity increase very fast as the
instability is approached due to contributions from values of $k$
close to $k_{c}$. For the same reason the decay of these
fluctuations becomes very slow. This is not the case for the
fluctuations of the other hydrodynamic fields, whose Langevin
equations are decoupled from Eq.\ ({\ref{4}}) \cite{LyL59}.
Therefore, it seems possible to consider a range of values of
$\widetilde{\delta L} \equiv (L_{c}-L)/L_{c}$ where the
fluctuations of ${\bm \omega}_{{\bm k},\perp}$, although still
small, dominate over the fluctuations of density and temperature.
But, although this is true for components with $k>0$, some care is
needed when analyzing Eq. (\ref{3}), since it involves the $k=0$
component of the temperature field, $\theta_{\bm 0}(s)$. The
Langevin equation for $\epsilon (s)$ is obtained from the
linearization around the HCS of the macroscopic average equation
for the total energy,
\begin{equation}
\label{7a}
\partial_{t}E(t)=-\frac{d}{2} \int d{\bm r}\, n({\bm r},t)
\zeta_{H}(n,T) T({\bm r},t).
\end{equation}
The result is:
\begin{equation}
\label{8}
\partial_{s} \epsilon (s)= \zeta^{*} \left[ \epsilon (s)-
\frac{3}{2V}\, \theta_{\bm 0}(s) \right].
\end{equation}
Here, the dependence of the cooling rate on the temperature has
been taken into account. Moreover, the noise term discussed in
\cite{BGMyR04}, associated with the localized character of the
energy dissipation, has been omitted. Although it can be expected
to be negligible far from the instability in the quasi-elastic
limit, this may be not the case near the instability. Equation
(\ref{8}) shows a coupling between the fluctuations of the volume
averaged temperature and those of the total energy. Use of
Eq.\ (\ref{3}) into Eq.\ (\ref{8}) and neglecting contributions
from the density and longitudinal velocity fluctuations gives
\begin{equation}
\label{9}
\partial_{s} \epsilon (s) = -\frac{\zeta^{*}}{2} \left[ \epsilon
(s)- \overline{\omega}(s) \right], \quad
\overline{\omega}(s)=\frac{6}{V^{2}d} \sum_{\bm k} |{\bm
\omega}_{{\bm k} \perp} (s)|^{2},
\end{equation}
valid in the region $\widetilde{\delta L} \ll 1$. The long time
limit of the average value of $\epsilon(s)$ is, therefore,
\begin{equation}
\label{10} \langle \epsilon \rangle_{st}= \lim_{s \rightarrow
\infty}
 \langle \overline{\omega}(s) \rangle = -\frac{3(d-1)}{Nd} \sum_{\bm k}
\frac{\eta^{*} k^{2}}{\lambda_{\perp}(k)}\, .
\end{equation}
Since we are considering $\widetilde{\delta L} \ll 1$, the sum
over ${\bm k}$ in the above expression is dominated by the $2d$
modes with the largest wavelength, for which
$\lambda_{\perp}(k_{min})\simeq -\zeta^{*} \widetilde{\delta L}$.
Using this into Eq. (\ref{10}), it follows that there is a
renormalization by fluctuations of the average total energy of the
HCS, $E(t)= \langle \tilde{E}(t) \rangle$, given by
\begin{equation}
\label{11} E(t) =E_{H}(t) \left[1+\frac{3(d-1)}{n_{H}L_{c}^{d}}\,
\widetilde{\delta L}^{-1} \right].
\end{equation}
Similarly, there is also a renormalization of the temperature of
the HCS, $T(t)=\langle \tilde{T}(t) \rangle_{st}$, that can be
evaluated directly from the long time limit of the average of Eq.\
(\ref{8}),
\begin{equation}
\label{12} T(t)= T_{H}(t) \left[ 1+\frac{\langle\theta_{\bm 0}
\rangle_{st}}{V} \right]=T_{H}(t) \left[ 1+\frac{2(d-1)}{n_{H}
L_{c}^{d}} \widetilde{\delta L}^{-1} \right].
\end{equation}
Alternatively, an effective temperature $T_{ef}(t)$ can be defined
as $T_{ef}(t)=2 E(t)/Nd$. Of course, the form of the renormalized
law for the temperature depends on the definition used for the
latter. In \cite{BGMyR05}, what was actually measured was
$\zeta_{ef}^{*}= \zeta_{ef}(T_{ef}) l_{0}/v_{H}(T_{ef})$, with
$\zeta_{ef}$ defined by $\partial_{t}T_{ef}=-\zeta_{ef}(T_{ef})
T_{ef}$. Then, it is found
\begin{equation}
\label{14}
\zeta^{*}_{ef}=\zeta^{*}\left[1+\frac{3(d-1)}{n_{H}L_{c}^{d}}
\widetilde{\delta L}^{-1} \right]^{-1/2}.
\end{equation}
This result predicts that near the clustering instability
threshold, $(\zeta^{*-2}_{ef}-\zeta^{*-2})/\zeta^{*-2}= A_{\zeta}
\widetilde{\delta L}^{-1}$ with $A_{\zeta}=3(d-1)/n_{H}L_{c}^{d}$,
that is just the behavior observed in \cite{BGMyR05}.

Define now $\widetilde{\delta E}(s)=[\tilde{E}(s)-E(s)]/E(s)=
[\epsilon(s)-\langle \epsilon\rangle_{st}] E_{H}(s)/E(s)$.
We are considering deviations from the
renormalized average value, i.e. including the fluctuations
effects, and not from the macroscopic bare value. A standard
calculation using Eq.\ (\ref{7}) and exploiting the Gaussian
character of the noise, gives that in the instability threshold
and for $s \geq s^{\prime} \gg 1$ it is
\begin{equation}
\label{15}
\langle[\overline{\omega}(s)- \langle \overline{\omega}
\rangle_{st}] [\overline{\omega}(s')- \langle \overline{\omega}
\rangle_{st}]\rangle_{st} =
\frac{9(d-1)}{n_{H}^{2}L_{c}^{2d} d}\,
\frac{e^{-(s-s^{\prime})/s_{c}}}{\widetilde{\delta L}^{2}} ,
\end{equation}
where $s_{c}=(2 \zeta^{*} \widetilde{\delta L})^{-1}$ is a
divergent ``critical'' relaxation time. Now Eq.\ (\ref{9}) can be
easily solved with the result, when $\widetilde{\delta L} \ll 1$,
\begin{equation}
\label{16} \langle \widetilde{\delta E} (s) \widetilde{\delta E}
(s^{\prime}) \rangle_{st}=
\langle[\overline{\omega}(s)- \langle \overline{\omega}
\rangle_{st}] [\overline{\omega}(s')- \langle \overline{\omega}
\rangle_{st}]\rangle_{st} ,
\end{equation}
valid for $s \geq s^{\prime} \gg 1$. Thus below the instability,
the scaled total energy fluctuations decay with the same rate as
the fluctuations of the kinetic energy associated with the
transversal modes of the velocity. For $s=s^{\prime}$, Eq.\
(\ref{16}) yields
\begin{equation}
\label{17} \sigma_{E}^{2} \equiv \langle (\widetilde{\delta
E})^{2} \rangle_{st}= A_{\epsilon}^{2} \widetilde{\delta L}^{-2},
\end{equation}
with $A_{\epsilon}^{2}=9(d-1)/n_{H}^{2} L_{c}^{2d}d$. Therefore,
close to the instability point, the relative dispersion of the
total energy fluctuations $\sigma_{E}$  presents a divergent
behavior with a critical exponent $-1$, and an amplitude
$A_{\epsilon}$ depending on the density $n_{H}$ and the
coefficient of normal restitution $\alpha$ (through the value of
the critical length $L_{c})$. Again, this is the same behavior as
reported in \cite{BGMyR05} from MD simulations.

To carry out a more detailed check of the theory presented here,
we have performed MD simulations of two-dimensional systems with
different values of $\alpha$ and $n_{H}$ (see Table \ref{table1}).
In all cases, the dependence on $\widetilde{\delta L}$ of both the
cooling rate and the dispersion of the total energy, i.e. the
exponents in the power laws (\ref{14}) and (\ref{17}), was in
agreement with the theoretical predictions. This was illustrated
in Figs. 1 and 2 of ref. \cite{BGMyR05} and no more details will
be given here. The comparison between the predicted critical
amplitudes and the MD results given in  Table \ref{table1} can be
considered as satisfactory, in the sense that the theory correctly
predicts the order of magnitude of the amplitudes, specially
taking into account the smallness of the quantities being
measured.

\begin{table}
\caption{Comparison between the predicted and MD values for the
critical amplitudes of the cooling rate, $A_{\zeta}$, and the
total energy dispersion, $A_{\epsilon}$. All the values of the
amplitudes have been multiplied by $10^{3}$.}
\begin{ruledtabular}
\begin{tabular}{cccccc}
$n_{H} \sigma^{2}$ & $\alpha$ & $ A_{\zeta}^{theory}$ &
$A_{\zeta}^{MD}$ & $A_{\epsilon}^{theory} $ & $A_{\epsilon}^{MD}$  \\
\hline 0.02 & 0.9 & 0.88 & 1.11 & 0.62 & 0.6 \\
0.02 & 0.8 & 1.62 & 3.63 & 1.15 & 1.5 \\
0.1 & 0.98 & 1.06 & 0.50 & 0.75 & 0.47 \\
0.1 & 0.95 & 2.4 & 2.38 & 1.7 & 1.45 \\
0.2 & 0.98 & 1.97 & 2.34 & 1.4 & 1.34 \\
0.2 & 0.95 & 4.59 & 7.78 & 3.24 & 3.6
\end{tabular}
\end{ruledtabular}
\label{table1}
\end{table}

\begin{figure}
\includegraphics[scale=0.65]{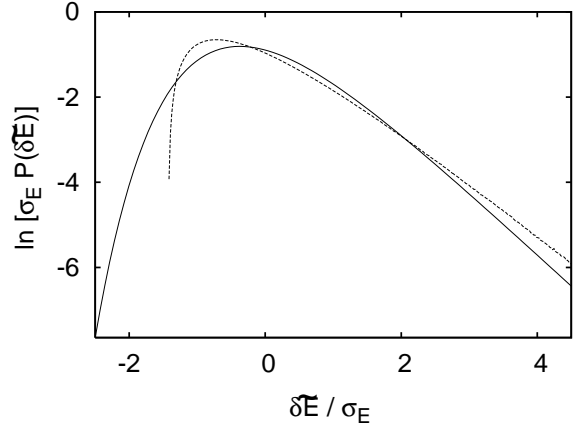}
\caption{Probability density function of the relative total energy
fluctuations $\sigma_{E} P_{L}(\widetilde{\delta E})$ for a system
of inelastic hard disks. The broken line is the theoretical
prediction derived in this paper and the solid line is Eq. (\ref{25}).
\label{fig1} }
\end{figure}

Next, let us proceed to investigate the form of the probability
distribution of the energy fluctuations. Particularization of Eq.\
(\ref{4}) for the modes with the smallest possible value of $k$ in
the limit $\widetilde{\delta L} \ll 1$ gives
\begin{equation}
\label{18} \left( \partial_{s} +\zeta^{*} \widetilde{\delta L}
\right) {\bm \omega}_{{\bm k} \perp}(s) = {\bm \xi}_{{\bm k}
\perp}(s),
\end{equation}
where it is understood that $|{\bm k}|=k_{min}$. Define a new time
scale $d\tau= \zeta^{*} \widetilde{\delta L}\; ds$, and a new
transversal velocity field by ${\bm \omega}^{*}_{{\bm k} \perp}=
{\bm \omega}_{{\bm k} \perp}/L_{c}^{d} \sigma_{E}^{1/2} $.
Equation (\ref{18}) becomes
\begin{equation}
\label{19} \left( \partial_{\tau} +1 \right){\bm \omega}^{*}_{{\bm
k} \perp} = {\bm \xi}^{*}_{{\bm k} \perp}(\tau),
\end{equation}
with
\begin{equation}
\label{20}
 \langle {\bm \xi}^{*}_{{\bm k} \perp}(\tau)  {\bm \xi}^{*}_{{\bm
k}^{\prime} \perp}(\tau^{\prime}) \rangle = \frac{d^{1/2}}{6
(d-1)^{1/2}}\,  \delta_{{\bm k},-{\bm k}^{\prime}} \delta (\tau
-\tau^{\prime}) {\sf I}.
\end{equation}
Equations (\ref{19}) and (\ref{20}) imply that the probability
distribution for
${\bm \omega}^{*}_{{\bm k} \perp}$, with $|{\bm k}|=k_{min}$, near
the clustering instability depends only on the dimension $d$ of
the system. In fact, since the noise term ${\bm \xi}^{*}_{{\bm k}
\perp}(\tau)$ is Gaussian, it is trivial to write the long time
form of this distribution using Eq.\ (\ref{7}) with
$s=s^{\prime}$,
\begin{equation}
\label{21} P_{st}({\bm \omega}^{*}_{{\bm k} \perp})= \left( 2 \pi
\sigma_{\omega}^{2} \right) ^{-(d-1)/2} e^{-  \omega^{*2}_{{\bm k}
\perp}/2 \sigma_{\omega}^{2}},
\end{equation}
with $\sigma_{\omega}^{2}= d^{1/2}/12 (d-1)^{1/2}$. In the time
scale $\tau$, and keeping only the dominant modes, Eq. (\ref{9})
reads
\begin{equation}
\label{22} \widetilde{\delta L}\,  \partial_{\tau} y =
-\frac{1}{2} \left( y- \frac{6}{d} \sum_{|{\bm k}|=k_{min}} |{\bm
\omega}^{*}_{{\bm k}\perp}(\tau)|^{2} \right),
\end{equation}
where $y= \epsilon/\sigma_{E}$ and the sum is restricted to
vectors ${\bm k}$ with $|{\bm k}|=k_{min}$. From the comparison
of Eqs.\ (\ref{19}) and (\ref{22}) it is seen that, on the $\tau$
scale and in the threshold of the instability, $y$ decays much
faster than the dominant components of ${\bm \omega}^{*}_{{\bm k}
\perp}$. Consequently, for large $\tau$ the solution of Eq.\
(\ref{22}) is $y= \frac{6}{d} \sum_{|{\bm k}|=k_{min}} |{\bm
\omega}^{*}_{{\bm k}\perp}(\tau)|^{2} $, where the probability
distributions of the modes ${\bm
\omega}^{*}_{{\bm k}\perp}$
is given by Eq. (\ref{21}).
Since the later does not depend on the parameters of the system
other than the dimensionality, the same property follows for the
probability distribution of both $y$ and the variable
\begin{equation}
\label{23} \frac{\widetilde{\delta
E}}{\sigma_{E}}=-[d(d-1)]^{1/2}+\frac{6}{d} \sum_{|{\bm
k}|=k_{min}} | {\bm \omega}^{*}_{{\bm k}\perp}|^{2}.
\end{equation}
This is equivalent to saying that the probability distribution for
$\widetilde{\delta E}$ verifies the scaling relation
\begin{equation}
\label{24} P(\widetilde{\delta E})= \frac{1}{\sigma_{E}} f \left(
\frac{\widetilde{\delta E}}{\sigma_{E}} \right),
\end{equation}
where $f$ is a scaling function. This is just the property assumed
in \cite{BGMyR05} and verified by MD simulations. Since the
probability distribution function for ${\bm \omega}^{*}_{{\bm k}
\perp}$ is known, it is possible to numerically generate the
probability distribution function $P(\widetilde{\delta E})$. The
result is shown  in Fig.\ \ref{fig1}. Also plotted is the function
\begin{equation}
\label{25} \Pi(\widetilde{\delta E})= K \left( e^{x-e^{x}}
\right)^{a}, \quad x=-b(s+\widetilde{\delta E}), \quad a=\pi/2,
\end{equation}
with $K=2.14$, $b=0.938$, and $s=0.374$, that fits extremely well
the MD results for $\sigma_{E}P(\widetilde{\delta E})$
\cite{BGMyR05}. It is important to remark that fluctuations in a
large number of equilibrium and non-equilibrium systems exhibiting
self-organized criticality as well as confined turbulent flows,
present the same kind of behavior \cite{BHyP98,BChFyal00}.
Although the agreement between both plotted curves is not so bad
for positive values of $\widetilde{\delta E}$, strong
discrepancies are observed for negative values. A major source for
them is easily identified from Eq.\ (\ref{23}), that for $d=2$
implies $\widetilde{\delta E}/ \sigma_{E} \geq - \sqrt{2}$, while
smaller values are found in the MD simulations. Since Eq.\
(\ref{23}) is a consequence of Eq. (\ref{8}), it seems plausible
that in order to elaborate a more accurate theory the intrinsic
noise associated with the cooling rate must be taken into account.

In summary, we have developed a mesoscopic theory for the
fluctuations of the total energy of an isolated granular gas near
the threshold of the clustering instability. The theory describes
accurately the qualitative behavior obtained in MD simulations,
namely the divergent behavior of the dimensionless second moment
and the decrease of the apparent cooling rate. Also, it is
consistent with the observed scaling property of the  probability
distribution function of the fluctuations. On the other hand, it
seems clear that a more refined formulation is needed in order to
get a more satisfactory quantitative agreement, especially for the
distribution function. In any case, we believe the present work
clearly indicates the way in which fluctuations in granular
systems near an instability can be analyzed and, in particular,
the dominant role played by nonlinear coupling between
hydrodynamic modes.

This research was supported by the Ministerio de Educaci\'{o}n y
Ciencia (Spain) through Grant No. FIS2005-01398 (partially
financed by FEDER funds).

\end{document}